\documentstyle[preprint,prb,aps]{revtex}

\tightenlines
\begin{document}

\title{Magnetic Polarons and the Metal-Semiconductor Transitions in (Eu,La)B$_6$ and EuO:  Raman Scattering Studies}
\author{C. S. Snow and S.L. Cooper}
\address{Department of Physics and Frederick Seitz Materials Research Laboratory, University of Illinois at Urbana-Champaign, 1110 West Green St., Urbana, IL 61801-9013}
\author{D.P. Young and Z. Fisk}
\address{National High Magnetic Field Laboratory, Florida State University, 1800 East Paul Dirac Drive, Tallahassee, Florida 32306, USA}
\author{Arnaud Comment and Jean-Philippe Ansermet}
\address{Ecole Polytechnique F\'ed\'erale de Lausanne, CH-1015 Lausanne, Switzerland}
\address{University of Illinois, Loomis Laboratory of Physics, 1110 West Green St., Urbana, IL 61801-9013}

\date{\today}
\maketitle

\begin{abstract}
We present inelastic light scattering measurements of EuO and Eu$_{1-x}$La$_{x}$B$_6$ ($x$=0, 0.005, 0.01, 0.03, and 0.05) as functions of doping, B isotope, magnetic field, and temperature.  Our results reveal a variety of distinct regimes as a function of decreasing T: (a)  a paramagnetic semimetal 
regime, which is characterized by a collision-dominated electronic scattering response whose scattering rate $\Gamma$ decreases with decreasing temperature; (b)  a spin-disorder scattering regime, which is characterized by a collision-dominated electronic scattering response whose scattering rate $\Gamma$ scales with the magnetic susceptibility; (c)  a magnetic polaron (MP) regime, in which the development of an $H$=0 spin-flip Raman response betrays the formation of magnetic polarons in a narrow temperature range above the Curie temperature T$_{\rm C}$; and (d)  a ferromagnetic metal regime, characterized by a flat electronic continuum response typical of other strongly correlated metals.  By exploring the behavior of the Raman responses in these various regimes in response to changing external parameters, we are able to investigate the evolution of charge and spin degrees of freedom through various transitions in these materials.  
\end{abstract}

\section{INTRODUCTION}
\label{sec:intro}

Strongly-coupled \hbox to 350pt{systems, i.e., materials having strong interactions among the electronic-,} \break spin-, and/or lattice-degrees of
freedom, have fascinated physicists for many years due both to their complex phase diagrams and to the diverse, exotic phenomena they exhibit.  Among the most intensely studied strongly-coupled systems, for example, are 
the high T$_{c}$ cuprates, which exhibit antiferromagnetic, ''pseudogap,'' and unconventional
superconducting phases, and the manganese perovskites, which have ''colossal
magnetoresistance''(CMR) paramagnetic, metallic ferromagnetic, and
charge-ordered antiferromagnetic phases.

One group of strongly-coupled materials that has been of particular interest recently is a broad class of doped magnetic semiconductors characterized by metal-semiconductor phase transitions, large negative magnetoresistance behavior near T$_{\rm C}$, and metallic ferromagnetism below T$_{\rm C}$.  One such system is EuO, which resistivity and magnetic measurements show has a ferromagnetic transition at 69 K, and an insulator-to-metal transition as the temperature is lowered below 50 K, resulting in a 13 orders-of-magnitude increase in the  conductivity.\cite{Torrance}  It has been proposed that the formation of magnetic (or ''spin'') polarons, i.e., ferromagnetic clusters of Eu$^{2+}$ spins, instigates this transition,\cite{Torrance} but thus far there has been no direct evidence for magnetic polaron formation in this system.  Another related low T$_{\rm C}$ metallic ferromagnet is EuB$_{6}$, which 
deHaas-van Alphen measurements\cite{GoodrichFS} and band structure calculations\cite{Massidda} 
indicate is an uncompensated semimetal at high temperatures.  Detailed magnetization and heat capacity measurements of EuB$_6$ reveal two transition temperatures at 
15.3 K and 12.5 K, associated respectively with
spin reorientation and ferromagnetic transitions;\cite
{SullowMag} transport,\cite{SullowMagPol} Raman,\cite{NyhusEuB6} and optical reflectivity
measurements\cite{DegiorgiRefl} further indicate that the transition near 12 K is associated with a semimetal to metal transition.  Raman scattering measurements reveal in this system the
spontaneous formation of magnetic polarons, again involving ferromagnetic clusters of Eu$
^{2+}$ magnetic moments, at temperatures just above T$_{\rm C}$.\cite{NyhusEuB6} The results of this study suggest that the magnetic polarons grow in size (and/or increase in number) with decreasing temperature or increasing magnetic field, implying that the paramagnetic semimetal (PMS) to ferromagnetic metal (FMM) transition in EuB$_6$ is also precipitated by the growth, and eventual percolation, of spin polarons.\cite{NyhusEuB6}  A recent transport and magnetization study of EuB$_6$ also suggests this interpretation.\cite{SullowMagPol}

Low carrier density magnetic systems such as EuB$_{6}$ (T$_{\rm C}$ $\sim$ 12 K)[ref. 4] and EuO (T$_{\rm C}$ $\sim$ 69 K)[ref. 1] are of interest 
because they share many properties with the CMR-phase manganese perovskites,\cite{NyhusEuB6,Yoon} including large negative magnetoresistance and magnetic cluster formation near T$_{\rm C}$, and complex metal-insulator transitions.  On the other hand, these Eu-based systems have few of the structural complexities of the manganites,\cite{Millis} such as strong electron-lattice effects associated with Jahn-Teller and breathing mode distortions.
EuB$_{6}$ is also interesting because its exotic
properties can be studied in a particularly controlled fashion, either as a function of doping via La substitution for
Eu (Eu$_{1-x}$La$_{x}$B$_{6}$), or as a function of isotopic substitution by replacing $^{10}$B for $^{11}$B. Therefore, (Eu,La)B$_{6}$ and EuO are ideal systems in which to investigate in a controlled manner a number of important issues and phenomena:  the effects of doping, disorder, isotopic substitution, temperature, and magnetic field on magnetic cluster formation;\cite{Moreo1,Calderon} metallic ferromagnetism; possible inhomogeneous magnetic phases; and the competition between ferromagnetic and antiferromagnetic correlations.
All of these subjects are pertinent to a wide class of itinerant ferromagnets such as La$_{1-x}$Ca$_{x}$CoO$_{3}$,\cite{Yamaguchi} the Mn perovskites (''manganites''),\cite{Moreo} tellurides like
Ge$_{1-x}$Mn$_{x}$Te and Eu$_{1-x}$Gd$_{x}$Te,\cite{Majumdar} and the Mn pyrochlores.\cite{Ramirez}

In this paper, we report a Raman scattering study of the complex magnetic phases of both EuO and Eu$_{1-x}$La$_x$B$_6$ as a function of doping ($x$), magnetic field, and temperature.  This study reveals a number of novel features of the phase diagram in these materials. For example, based upon 
the development of a zero-field spin-flip Raman response above T$_{\rm C}$ in both EuO and Eu$_{1-x}$La$_x$B$_6$, we conclude that magnetic polarons form in a narrow range above T$_{\rm C}$ in these materials, and that these polarons evolve continuously through their respective semiconductor/(semimetal)-metal transitions into a ferromagnetic metal phase. As we will show, isotope effect studies provide evidence that these polarons are predominantly magnetic, with a negligible lattice contribution. Moreover, a comparison of magnetic polaron formation in EuO and (Eu,La)B$_6$ provides evidence that the lower carrier mobility in the former system plays an important role in stabilizing magnetic polarons, both above and below T$_{\rm C}$. We note that magnetic polarons have been observed previously by Raman scattering in the paramagnetic phase of dilute magnetic semiconductors (DMS) such as CdMnTe [ref. 17] and Cd$_{0.95}$Mn$_{0.05}$Se:In.\cite{Dietl} However, the Raman studies of EuO and (Eu,La)B$_6$ presented in this study reveal new details of magnetic polaron formation and its relationship to metal-semiconductor transitions in complex magnetic systems:  For example, EuO and (Eu,La)B$_6$ have phase diagrams akin to those of strongly-correlated systems such as the CMR manganites, and consequently, one can explore the properties and evolution of magnetic polarons in very different phase regions than has been possible in DMS systems. Additionally, due to the rather high transition temperatures of EuB$_6$ (T$_{\rm C}$ $\sim$ 12 K) and EuO  (T$_{\rm C}$ $\sim$ 69 K), Raman studies of these materials afford an opportunity to investigate the evolution of magnetic polarons through exotic phase transitions, particularly the important paramagnetic semiconductor/(semimetal) to ferromagnetic metal transition, which has not been possible in DMS.
  
The remainder of this paper is organized as follows:  Section II describes the experimental techniques used, and sample preparation and characterization methods; and Section III discusses the temperature dependence, isotope effect, magnetic field dependence, and doping dependence of the Raman scattering spectra in the various magnetic phases of (Eu,La)B$_6$ and EuO, including the spin-fluctuation dominated PM regime (Section III.A), the magnetic polaron regime (Section III.B), and the ferromagnetic metal regime (Section III.C).

\section{EXPERIMENTAL}
The principal measurement technique used in this study is Raman scattering, which has proven to be a powerful tool for studying strongly-correlated systems for a number of reasons.  First, experimental investigations of strongly-coupled systems are complicated by the fact that phase changes in these compounds generally involve simultaneous changes in the charge-, spin-, and lattice-degrees of freedom.  While most experimental probes are primarily useful for probing either charge or spin excitations, Raman scattering has proven to be an effective tool for studying the complex phase changes exhibited by strongly-coupled systems because it can simultaneously convey energy, lifetime, and symmetry information about lattice, electronic, as well as magnetic excitations.  This has been illustrated for example in studies of hexaborides,\cite{NyhusEuB6,NyhusSmB6} manganites,\cite{Yoon} ruthenates,\cite{HLLiu} and nickelates.\cite{Yamamoto}  
Moreover, Raman measurements are particularly sensitive to subtleties in the phase diagram of strongly-coupled systems, for example revealing spin-disorder, bound magnetic polaron, and ferromagnetic metal phase regimes in low-carrier density magnetic systems;\cite{NyhusEuB6} consequently, this technique provides an ideal method for exploring the sensitivities of these phases to doping, temperature, and magnetic field.  Finally, Raman scattering is particularly beneficial as a probe of strongly-coupled systems because it lends itself to specialized studies that are difficult using other techniques, such as high pressure, high magnetic field, and high spatial resolution measurements.

\subsection{Techniques and Equipment}
All of the Raman scattering measurements in this study were performed in a true backscattering configuration on the (100) surfaces of cubic EuO $\lbrack$$O_h$$^5$ - $Fm\bar{3}m$$\rbrack$ and cubic Eu$_{1-x}$La$_{x}$B$_{6}$ ($x$=0,0.005,0.01,0.03,0.05) $\lbrack$$O$$_{h}$$^{1}$$-Pm3m$$\rbrack$ single crystal samples grown in an aluminum flux. The spectra were analyzed using a modified subtractive triple stage spectrometer and measured with a liquid-nitrogen-cooled Photometrics CCD. The temperature and magnetic field were varied by placing the sample in a pumped Oxford flow cryostat, which was mounted in the bore of an Oxford superconducting magnet.
Temperatures as low as 4 K, and magnetic fields as high as 8 Tesla, were
obtained. The samples were excited with the 6471 {\AA } line from a Krypton
ion laser with an incident power of 5mW focussed to a $\approx$50 $\mu$m spot. Spectra were obtained with the incident and scattered light
polarized in the following configurations in order to identify the
symmetries of the excitations studied:  ({\bf E$_{i}$, E$_{s}$})=({\bf x,x}):  A$%
_{1g}$+E$_{g}$; ({\bf E$_{i}$, E$_{s}$})=({\bf x,y}):  T$_{2g}$+T$_{1g}$; ({\bf %
E$_{i}$, E$_{s}$}) =({\bf x+y,x+y}):  A$_{1g}$+$\frac{1}{4}$E$_{g}$+T$_{2g}$; (%
{\bf E$_{i}$, E$_{s}$}) =({\bf x+y,x-y}):  $\frac{3}{4}$E$_{g}$+T$_{1g}$, where A$_{1g}$, E$_g$, and T$_{1g}$/T$_{2g}$ are respectively the single, doubly, and triply degenerate irreducible representations of the $O_h$ space group.  In magnetic 
field experiments, it was necessary to use circularly polarized light in order to avoid symmetry mixing due to 
Faraday rotation effects.  Circularly polarized light was produced with a
polarizing cube and a Berek compensator optimized for the 6471{\AA } line of
the Krypton ion laser. The scattered light was analyzed with a broadband
quarter wave plate and a polarizing cube. The magnetic field spectra were measured in the ({\bf E$_{i}$},{\bf E$_{s}$})=({L},{L}): A$_{1g}$($\Gamma $$_{1}$$^{+}$) + $1 \over 4$E$_g$($\Gamma$$_{3}$$^{+}$) + T$_{1g}$($%
\Gamma $$_{4}$$^{+}$) and ({\bf E$_{i}$},{\bf E$_{s}$})=({L},{R}): $3 \over 4$E$_g$($\Gamma$$_{3}$$^{+}$) + T$_{2g}$($%
\Gamma $$_{5}$$^{+}$) symmetries, where R(L) stands for right (left) circular polarization.

Optical reflectance measurements were also performed between 100 cm$^{-1}$ and 50000 cm$^{-1}$ (12 meV $-$ 6.2 eV), and were obtained with a Bomem rapid scanning interferometer
in a near-normal incidence configuration. The modulated light beam from the
interferometer was focused onto either the sample or an Au reference mirror,
and the reflected beam was directed onto a detector appropriate for the
frequency range studied. The different sources, polarizers, and detectors
used in these studies provided substantial spectral overlap, and the
reflectance mismatch between adjacent spectral ranges was less than 1{\%}.

Magnetic susceptibility measurements were taken on a Quantum Design 5 Tesla MPMS SQUID Magnetometer.  The samples were zero-field-cooled to $\approx$2 K and then measured as a function of temperature in a constant field of 1000 Gauss.  No demagnetization corrections have been made to the data.
\subsection{Materials}
The substitution of La$^{3+}$ for Eu$^{2+}$ adds one free electron per La, causing profound changes in the phase diagram and transport properties of Eu$_{1-x}$La$_{x}$B$_{6}$.  A study of this system therefore allows a detailed study of the effects of doping on magnetic polaron formation, the 
metal-semimetal transition, and
the competition between ferromagnetism and antiferromagnetism in complex magnetic systems.

The effects of doping on the transport and magnetic properties of EuB$_6$ can be illustrated by examining the magnetic susceptibility, resistivity, and optical response as a function of La$^{3+}$ substitution.  Figure~\ref{susc resist} (a) shows the magnetic susceptibility data as a function of temperature and doping: at low temperatures, the magnetic susceptibility 
decreases systematically from $x$=0 to $x$=0.05. Figure~\ref{susc resist} (b) shows the resistivity data as a function of temperature and doping, illustrating an obvious difference in behavior between $x$ $\leq $0.01 and $x$ $\geq $0.03 doping regimes at low temperatures.  Specifically, for $x$ $\leq $0.01, the resistivity drops at temperatures below 15 K as the materials transition into the ferromagnetic-metal 
ground state, while for $x$ $\geq $0.03, the low temperature resistivity data exhibit a sharp increase below 10 K.

Figure~\ref{optics} shows the room-temperature optical conductivity ($\sigma $$_{1}$($\omega $)) [Fig.~\ref{optics} (a)] and loss
function (Im($\frac{-1}{\epsilon }$)) [Fig.~\ref{optics} (b)] as a function of doping, obtained from a Kramers-Kronig
transformation of the measured optical reflectivity.  Throughout the La substitution range, the optical conductivity
of Eu$_{1-x}$La$_x$B$_6$ exhibits a classic Drude response at T=300 K, and the loss function exhibits a clear 
plasmon response peaked at 1195, 1205, 1235, 2027, and 2221 cm$^{-1}$ for $x$=0, $x$=0.005, $x$=0.01, $x$=0.03, and $x$=0.05, respectively.  The extremely low plasmon energy in these systems is consistent with low carrier density semimetallic behavior, and the trend exhibited by the plasmon energy with doping is consistent with the assumption that La substitution adds carriers to the EuB$_6$ system.  

Interestingly, between $x$=0.01 and $x$=0.03 there is an abrupt increase in the plasmon response, and a dramatic change in behavior in the resistivity, 
magnetic susceptibility, and optical conductivity data that can be attributed to one of two likely sources: (1)  There is an uncertainty in the actual value of La doping; and (2) a peak in the DOS, predicted in the band structure calculations of Massidda {\it et al.},\cite{Massidda} causes a dramatically increased metallic behavior in the $x$=0.03 sample compared to the $x$=0.01 sample.  In the latter case, the abrupt change in behavior with doping would likely be observed in all doped divalent hexaborides.  

These data, and the data to be presented later in this paper, is consistent with  a {\it qualitative} phase diagram for the (Eu,La)B$_{6}$ system presented in Figure~\ref{phasediagram}; notably, this phase diagram is similar to that
inferred from neutron scattering studies on EuB$_{6-x}$C$_{x}$, in which
doping via C substitution (with increasing $x$) induces a phase change from a
ferromagnetic state, to a mixed ferromagnetic/antiferromagnetic phase, and
finally to an antiferromagnetic state.\cite{Tarascon}  Another interesting
parallel is to the Eu chalcogenides, which at low temperatures exhibit a systematic change, as a function of increasing radius of the chalcogenide atom, from metallic ferromagnetism in EuO, to a metamagnetic structure in EuSe, and finally to antiferromagnetism in EuTe.\cite{Leroux-Hugon} 

\section{Results and Discussions}
\subsection{Electronic Raman Scattering In The Paramagnetic Phase}
The Raman scattering spectrum of Eu$^{11}$B$_6$ in the PM phase is illustrated for the frequency range 0 - 1500 cm$^{-1}$ in Fig.~\ref{EuB6 300K}, showing: (i) three distinct optical modes involving the boron atoms at 762 cm$^{-1}$ (T$_{2g}$), 1098 cm$^{-1}$ (E$_g$), and 1238 cm$^{-1}$ (A$_{1g}$); and (ii) a ''diffusive'' electronic Raman scattering response, i.e., S($\omega$) $\sim$ 1/ $\omega$ (Eq. (2)), at low frequencies ($<$ 500 cm$^{-1}$).  In the remainder of this paper, we will primarily focus on the behavior of electronic and spin-flip Raman scattering in this low frequency ($<$ 500 cm$^{-1}$) region. 

To understand the origin of the low-frequency electronic Raman scattering response seen in the PM phase of EuB$_6$ (and EuO), we note first that when the electronic scattering rate is less than the product of the wave vector and Fermi velocity ($\Gamma < qv_F$), the electronic Raman 
scattering response reflects the density-density correlation function, and has a differential scattering cross section given by\cite{Klein}    
\begin{equation}
{{d^2 \sigma} \over {d \omega d \Omega}} \propto (1+n(\omega)) r^{2}_0 (\hat\epsilon_i \cdot \hat\epsilon_s)^2
{{\hbar k^2} \over {4 \pi^2 e^2}} Im \bigl\lbrack{{-1} \over {\epsilon_0 (\vec{k},\omega)} }\bigr\rbrack
\end{equation}
where $r_0$ is the classical radius of an electron (=${e^2 \over {mc^2}}$), 
$\hat\epsilon_i$ and $\hat\epsilon_s$ are the incident and scattered light 
polarization vectors, respectively, $\vec{k}$ is the wave vector, $1+n(\omega)$ is the Bose-Einstein thermal factor, and $\epsilon_0$ is 
the dielectric response function.  Therefore, at high frequencies, the electronic scattering response predicted by Eq. 1 is proportional to the dielectric loss function, and is consequently characterized by a peak in the Raman scattering response near the plasmon energy.  At low frequencies, one expects in this regime a negligible electronic scattering response, due to the fact that Coulomb correlations push most of the low-frequency scattering strength in Eq. 1 into the plasmon response.\cite{Klein}

On the other hand, when the carrier scattering rate is greater than $qv_F$, i.e., $\Gamma > qv_F$, for example due to strong electronic scattering from spins, impurities, phonons, etc., then the low frequency Raman intensity from mobile carriers can become appreciable, and is expected to exhibit a ''collision-dominated'' scattering response, 
\cite{Ipatova}
\begin{equation}
{{d^2 \sigma} \over {d \omega d \Omega}} \propto (1+n(\omega))\mid \gamma_L \mid^2 
\frac{\omega \Gamma_L }{\omega^{2}+ \Gamma_L^{2}}
\end{equation}
where $\gamma_L$ is the Raman scattering vertex associated with scattering geometry $L$, and $\Gamma_L$ is the carrier scattering rate associated with scattering geometry $L$.  Such collision-dominated scattering is clearly observed at low frequencies in EuB$_6$ (see Fig.~\ref{EuB6 300K}), in the low $x$ Eu$_{1-x}$La$_x$B$_6$ compounds (see Fig.~\ref{300k}), and in EuO (see Fig.~\ref{EuO 1}).  Note that the collision-dominated Raman scattering response is the analogue of the Drude optical response due to free carriers, and hence the carrier dynamics of a material can be effectively studied with light scattering in the $\Gamma > qv_F$ regime. 

Interestingly, Fig.~\ref{300k} shows that the ''collision-dominated'' low-frequency electronic scattering response in Eu$_{1-x}$La$_x$B$_6$ is gradually suppressed as a function of increased La concentration, and is very weak in the highly-doped Eu$_{0.95}$La$_{0.05}$B$_6$ sample. Importantly, this suppression of low frequency electronic scattering intensity cannot be attributed to increased carrier localization, as Fig.~\ref{optics} (b) and the inset of Fig.~\ref{300k} clearly illustrate that there is an {\it increase} in the carrier density with increasing La substitution. Rather, the suppression of scattering with increasing $x$ most likely arises from the fact that in higher carrier density systems, i.e., those in which $\Gamma < qv_F$, one expects the electronic scattering response in Eq. 1 to have negligible weight at low frequencies, due to the fact that Coulomb correlations push most of the low frequency scattering strength into the high frequency plasmon response.\cite{fermi liquid}

Typically, the scattering rate $\Gamma$ described in Eq. 2 reflects carrier scattering from static impurities.\cite{Ipatova} Importantly, however, in doped ferromagnetic systems such as (Eu,La)B$_6$ and EuO, the carrier scattering rate $\Gamma$ in Eq. 2 should be dominated by spin-fluctuation scattering over a substantial temperature range above T$_{\rm C}$; consequently, the associated collision-dominated scattering response in this temperature regime should afford direct information regarding the spin dynamics near T$_{\rm C}$.  In particular, in low carrier density magnetic semiconductors/semimetals, the collision-dominated scattering rate in this temperature regime should satisfy the relationship $\Gamma \propto  l \propto T \chi$,\cite{Leroux-Hugon,DietlSpalek} where $l$ is the electronic mean free path, $\Gamma$ is the carrier scattering rate, T is the temperature, and $\chi$ is the 
magnetic susceptibility.  Indeed, as shown in Fig.~\ref{gamma chi}, a scaling between T$\chi$ and the low-frequency electronic Raman scattering rate $\Gamma$, estimated from fits of the data by Eq. 2, is clearly observed in EuB$_6$ within the temperature range 35 K to 65 K, providing strong evidence that spin-fluctuations are the principal scattering mechanism for the carriers in this temperature range.  Similar 
''spin-fluctuation-induced'' electronic Raman scattering is also observed in the paramagnetic phase of EuO above T$_{\rm C}$: Fig.~\ref{EuO 1} (b) illustrates that the integrated intensity of the collision-dominated scattering response (Eq. 2) in the PM phase of EuO diverges for T $\rightarrow$ T$_{\rm C}$, again due to the enhancement of carrier scattering by spin fluctuations (note also the development of an inelastic peak near T$_{\rm C}$ in Fig.~\ref{EuO 1} (a), which we attribute to zero-field spin-flip Raman scattering associated with the development of magnetic polarons; see Section III.B).   

These results confirm that for a substantial temperature range above T$_{\rm C}$, the carrier dynamics of both EuO and (Eu,La)B$_6$ are dominated by spin-disorder scattering over a wide temperature range above T$_{\rm C}$ in both these systems.  Importantly, EuO exhibits spin-fluctuation induced electronic Raman scattering up to much higher temperatures than (Eu,La)B$_6$, indicating a much greater prevalance of spin disorder in the former compound.  These results also illustrate that electronic Raman scattering can probe spin dynamics and magnetotransport in doped magnetic systems, suggesting a means by which magnetotransport can be explored under various extreme conditions that may not be easily investigated using other techniques, such as  at ultra-high pressures ($>$ 20 GPa), with high ($\sim$ 1 micron) spatial resolution, and in electronically inhomogeneous systems.\cite{Liu magneto} 

\subsection{Magnetic Polaron Regime} 
\subsubsection{Temperature Dependence at H=0:  EuB$_6$ and EuO}
Figs.~\ref{bmp eub6} and~\ref{bmp euo} (a) show that in a narrow temperature range just above T$_{\rm C}$ (for $H$=0), the collision-dominated electronic scattering response of both EuB$_6$ and EuO gradually disappears with decreasing temperature, and a broad, gaussian-shaped inelastic peak develops between 50-100 cm$^{-1}$ in the {\bf E$_i$} $\bot$ {\bf E$_s$} scattering geometry (see Fig.~\ref{bmp euo} (b)).  As shown previously in numerous Raman scattering studies of dilute magnetic semiconductors, the presence of this distinctive Raman scattering response has been identified with zero-field spin-flip (SF) Raman scattering associated with the presence of magnetic polarons.\cite{Dietl,magpolrefs,Isaacs,Heiman} Consequently, we infer from the development of a zero-field SF Raman peak in both EuB$_6$ and EuO that magnetic polarons develop in a narrow temperature range above T$_{\rm C}$ in both these systems. This evidence for magnetic polaron formation above T$_{\rm C}$ in EuB$_6$ and EuO is consistent with earlier reports of magnetic polaron formation in EuB$_6$, as inferred from Raman scattering\cite{NyhusEuB6} and transport\cite{SullowMagPol} measurements, and in EuO, as inferred from transport and magnetic measurements.\cite{Torrance,Mauger}

There are several important points to make regarding evidence for magnetic polaron formation above T$_{\rm C}$ in both EuB$_6$ and EuO:  First, magnetic polaron development in these materials is favored by their large ferromagnetic {\it d-f} exchange couplings (J$_{df} \sim$ 0.1 eV), which causes a trapped charge to lower its energy by polarizing the local moments (Eu$^{2+}$ spins) inside its Bohr radius. The resulting exchange-field associated with these polarized moments causes a splitting of the carrier's spin-up and -down levels, which is observed in Raman scattering measurements through the appearance of a $H$=0 SF Raman peak.  The resulting zero-field Raman shift associated with this spin-flip process is expected to be proportional to the {\it d-f} exchange-induced spontaneous magnetization of the local Eu$^{2+}$ moments within the magnetic polaron, and it is associated with an effective exchange field within the polaron,\cite{Isaacs}
\begin{equation}
H_{ex} = {{\hbar \omega_o} \over {g \mu_B}}
\end{equation}
where $g$ is the Land\'e $g$ factor, and $\mu_B$ is the Bohr magneton.  Note that the large energy shifts (50-100 cm$^{-1}$) observed for the zero-field SF Raman energies in EuB$_6$ and EuO are much larger than typical Zeeman energies, reflecting the large values of the {\it d-f} exchange-induced magnetization and exchange-field $H_{ex}$ in these systems.  

Second, the zero-field SF Raman contribution to the spectrum (obtained after subtracting a small diffusive contribution [eq. 2], as shown in Fig.~\ref{bmp euo} (a)), is well described by a gaussian lineshape,
\begin{equation}
S(\omega) \sim \omega_o^2exp \lbrack({\hbar \omega_o}^2)/{2 W^2}\rbrack 
\end{equation}
as is typical of zero-field SF Raman scattering spectra associated with magnetic polarons in dilute magnetic 
semiconductors.\cite{Isaacs,Heiman}  In Eq. 4, the peak frequency, $\hbar \omega_o$, represents the spin-splitting of the trapped charge due to the local exchange-field of the Eu$^{2+}$ moments within its Bohr radius, and the gaussian linewidth $W$ represents a  probability distribution of the number of magnetic ions contributing to the magnetization within different Bohr orbits, due primarily to thermal and compositional fluctuations.\cite{Isaacs,Heiman}  Importantly, the inset of Fig.~\ref{bmp euo} (a) illustrates that the SF peak frequency $\hbar \omega_o$ of EuO increases systematically with decreasing temperature.  This is indicative 
of magnetic polarons in a spin-aligned, ''cooperative'' regime wherein the carrier spin and the magnetic ion spins are mutually aligned in the ferromagnetic clusters.\cite{Isaacs,Heiman}  This behavior is contrasted with the temperature-independent peak frequency expected in the ''spin-fluctuation'' regime, in which the carrier spin interacts with a non-zero local magnetization associated with random fluctuations of the moments, but in which there is no correlation among the Eu$^{2+}$ spins.\cite{Isaacs} Finally, it should be noted that, in contrast to dilute magnetic semiconductor systems such as Cd$_{1-x}$Mn$_x$(Te,Se,S), the temperature dependence of the zero-field SF Raman response is quite rich in the ''dense'' EuB$_6$ and EuO systems, due in part to the presence of complex metal-insulator and magnetic phase changes in the Eu$^{2+}$ based systems. Indeed, the zero-field SF Raman scattering response in both EuB$_6$ and EuO disappears above a temperature we identify as the magnetic polaron formation temperature T$_p$ (see Figs.~\ref{BMP} and~\ref{BMPsumm}), due to a reduction in the magnetic susceptibility and to the disruption of spin alignment in the clusters by thermal fluctuations.  Moreover, the zero-field SF Raman response diminishes in the ferromagnetic metal phase [T$<$ T$_{\rm C}$] (see Figs.~\ref{bmp eub6} and~\ref{bmp euo}), due to the ferromagnetic alignment of large portions of the material, and the resulting delocalization of many of the bound charges.

The observation of spin-flip Raman scattering responses above T$_{\rm C}$ and at $H$=0 in EuB$_6$ (Fig.~\ref{bmp eub6}) and EuO (Fig.~\ref{bmp euo} (a)) clearly demonstrate that the semiconductor-metal transitions of both these materials are preceded by the development of ferromagnetic clusters in the PM phase.  Importantly, the spectral response associated with spin-flip Raman scattering also affords specific information regarding the magnetic polarons in these materials.  Most theoretical studies of magnetic polarons have been confined to dilute magnetic semiconductors such as Cd$_{1-x}$Mn$_x$Te,\cite{Nawrocki,DietlSpalek,Umehara} which do not exhibit transitions into a FM metal phase.  Consequently, we expect that while these theoretical results might be generally applicable to our measurements in the PM phase, they will likely have the wrong energy scales, and will not address the physics of the percolation transition into the FM metal phase, in systems such as EuB$_6$ and EuO.  However, the stability criterion for magnetic polarons in magnetic semiconductors exhibiting MI transitions into a FM phase has been treated in several limiting cases.\cite{Kasuya,MaugerKublerVigren}  For example, in the limiting case of large {\it d-f} exchange, Kasuya {\it et al.} [ref. 33] found that a well-defined small paramagnetic polaron will form at a temperature $T_p$ above T$_{\rm C}$, with a radius given by
\begin{equation}
R = {{2E_0} \overwithdelims () {3 \eta k_B {\rm T_p}}}^{1/5}a
\end {equation}
where T$_p$ is the polaron stabilization temperature, $E_0$ is the conduction band width, $a$ the lattice parameter, and  
\begin{equation}
\eta = ln(2S+1) - {{3S {\rm T_{\rm C}}} \over {2(S+1){\rm T_p}}}.
\end{equation}
For EuB$_6$, using measured and known values of T$_{\rm C}$ = 12 K, $E_0$ = 0.25 eV,\cite{Massidda} T$_p$ = 30 K, and S = 7/2 in the Kasuya model, we obtain an estimate of $R \sim$ 2$a$;  similarly, for EuO, using T$_{\rm C}$ = 69 K, $E_0$ = 1 eV,\cite{Leroux-Hugon} T$_p$ = 90 K, and S = 7/2, we obtain an estimate of $R \sim$ 2.5$a$.  Importantly, the Kasuya model not only significantly overestimates our observed polaron stabilization temperatures in EuB$_6$ and EuO, but also likely underestimates the size of $R$ in these systems.  These errors are due primarily to two factors:\cite{droplets} (i)  uncertainty in the known experimental parameters; and (ii)  approximations in the Kasuya model, particularly the failure to consider thermodynamic fluctuations in the magnetization, which have been shown to make an important contribution to polaron stabilization in dilute magnetic semiconductors.\cite{DietlSpalek,Umehara,MaugerKublerVigren}  

Other pertinent studies of magnetic polaron formation in magnetic semiconductors have included a Monte-Carlo simulation,\cite{Calderon} and a variational calculation that solved the Schroedinger equation for spin configurations of up to 13 spins.\cite{Guillaume}  While the results of these calculations do not allow a quantitative comparison to our experimental data, the qualitative predictions of these calculations are borne out in our experimental results.  In particular, the variational calculation\cite{Guillaume} indicates that the distribution of polaron sizes, the average polaron size, and the average polaron energy all increase with decreasing temperature towards T$_{\rm C}$.  These trends are consistent with those observed for the spin-flip Raman energies in (Eu,La)B$_6$ and EuO, as summarized in Figs.~\ref{BMP} and \ref{BMPsumm}.    

Interestingly, a comparison of the properties and energy scales of the magnetic polarons in EuB$_6$ and EuO, as inferred from our $H$=0 SF Raman results, illustrate several basic differences between these two materials.  First, the magnetic polarons in EuO are stable over a much wider temperature range than in EuB$_6$, as is clearly shown by comparing the quantity (T$_p$ - T$_{\rm C}$) observed for these two systems, where (T$_p$ - T$_{\rm C}$) is the difference between the polaron formation temperature and the Curie temperature.  In EuB$_6$, (T$_p$ - T$_{\rm C}$) is approximately 2/3 that of EuO, i.e., 15 K compared to 20 K.  The increased polaron stability in EuO compared to EuB$_6$ can be understood in part by considering the very different carrier mobilities exhibited by these two materials in the PM phase:  EuO exhibits an activated hopping conductivity (E$_A \sim$ 0.3 eV [ref. 1]), while EuB$_6$ has a metallic conductivity and a well-defined plasma edge; consequently, the lower carrier mobilities in EuO allow interactions affiliated with carrier localization to compete more effectively with the carrier kinetic energy throughout a substantially broader temperature range in EuO ($\sim$ 20 K [ref. 1]) than in EuB$_6$ ($\sim$ 2 K [ref. 4]).  

It is also interesting to note that the energy of the polaron in EuO is approximately 2/3 the energy of the polaron in EuB$_6$.  
Assuming comparable values of the exchange coupling ($\sim$ 0.1 eV), and the number of Eu$^{2+}$ moments contributing to the magnetization in the polaron, $\bar{x}$, this difference appears to reflect a greater alignment of spins in the magnetic polarons of EuB$_6$ than EuO.  This, in turn, would be indicative of a larger amount of spin-disorder in EuO than EuB$_6$, of which there are two possible sources:  First, the higher resistivities in EuO suggest that there is likely to be more static spin disorder in this system than in EuB$_6$.  Indeed, the presence of substantial static spin-disorder in EuO is supported by the fact that magnetic polarons appear to persist even well below T$_{\rm C}$ in EuO [see Fig.~\ref{bmp euo} (a)], but disappear abruptly below T$_{\rm C}$ in EuB$_6$ [see Fig.~\ref{bmp eub6}];  see Section III.C.  Second, since EuO has a higher polaron formation temperature (T$_p$) than EuB$_6$, thermal fluctuations will have a greater impact on the net magnetization of the polarons in EuO.  

\subsubsection{Isotope Effect:  Eu$^{11}$B$_6$ vs. Eu$^{10}$B$_6$} 
It is also of interest to examine the effect of boron isotope on the SF Raman energy in EuB$_6$ in order to study the possible effects of the lattice on spin polaron formation and the PM semiconductor - FM metal transition in this material.  Fig.~\ref{boron10} shows that the three boron optical modes exhibit a clear shift of 5$\%$ upon the substitution of $^{10}$B for $^{11}$B in EuB$_6$, consistent with the mass ratio of $\sqrt{\frac{11}{{10}}}$.  By contrast, however, no significant difference is detected between the SF Raman response observed in Eu$^{11}$B$_{6}$ and Eu$^{10}$B$_{6}$ (see inset to Fig.~\ref{boron10}): both samples undergo PS to FM transitions at the same temperature, and the SF Raman energies are the same in both samples.  This seems to suggest a negligible lattice component in the magnetic polarons formed in EuB$_6$.  However, it should be noted that shifts of only 5$\%$ are difficult to unambiguously identify, due to the width of the SF Raman peak and the approximately 2 cm$^{-1}$ resolution of the spectrometer.  Nevertheless, while it is impossible to say with certainty that magnetic polarons in EuB$_6$ do not couple to the lattice degrees of freedom, these data suggest that any such coupling is small.  Consequently, the polarons in EuB$_6$ and probably EuO are most likely pure spin polarons with negligible lattice contributions.  This reflects a distinct difference between magnetic polarons formed in low carrier density systems such as EuB$_6$ and EuO, which can be stabilized solely by static disorder and the large ferromagnetic exchange interaction in these systems, and {\it magnetoelastic} polarons that form in the high-carrier-density manganites, which demand large Jahn-Teller and breathing mode distortions to help balance much larger carrier kinetic energies.\cite{Millis}

\subsubsection{Magnetic Field Effects}
The magnetic field dependence of the SF Raman scattering responses in EuB$_6
$ and EuO is also quite interesting.  Figs.~\ref{BMP} and~\ref{BMPsumm} illustrate several significant trends associated with the SF Raman response as a function of magnetic field.  First, the SF Raman energy increases systematically with increasing magnetic field, suggesting that there is an increase in the exchange energy of the magnetic polarons with increasing field.  The large increase in the SF Raman energy with field in both EuB$_6$ and EuO (see Figs.~\ref{BMP} and~\ref{BMPsumm}) is caused by the fact that the magnetization of the Eu$^{2+}$ ions enhance the effect of the magnetic field on the spin-splitting of the bound charge.  This enhancement is reflected in a {\it magnetic field-induced contribution} ({\it i.e.}, in addition to the $H$=0 SF contribution) to the spin splitting of the bound 
charge given by\cite{magpolrefs}
\begin{equation}
\Delta (H) \sim \bar{x} \alpha N_o({7/2})B_{7/2}({{g \mu_B H} \over {k_B T^*}}) + g^* \mu_B H, 
\end{equation}
where $\bar{x}$ is the number of Eu$^{2+}$ moments contributing to the magnetization, $\alpha N_0$ is the exchange constant, $B_{7/2}$ is the Brillouin function for J=7/2, {\it g} is the Land\'e factor, g$^*$ is the intrinsic {\it g}-factor,\cite{Dietl} and T$^*$ is an effective temperature that appropriately accounts for magnetic correlations.\cite{t-tc}  
Eq. 7 predicts that $\Delta$ should vary linearly with magnetic field $H$ for small $H$, then saturate for large $H$, consistent with our observations (see Fig.~\ref{BMPsumm}).  Interestingly, in its linear-in-field regime, Eq. 7 can be written as\cite{Isaacs,Heiman}
\begin{equation}
\Delta (H) \sim g_{eff} \mu_B H
\end{equation}
where
\begin{equation}
g_{eff} = {1 \over \mu_B}({{\delta \Delta} \over {\delta H}}) = (3/2)(\bar{x} \alpha N_o g /k_B T^*) + g^*
\end{equation}
is an effective $g$-factor that reflects the amplified field-induced splitting of the carrier levels due to the {\it d-f} exchange interaction.  
Note from Eq. 9 that one can obtain an experimental estimate of $g_{eff}$ from the measured slope of $\hbar \omega_o$ vs. $H$ in the linear-in-field regime. In our results, such an estimate is complicated by the fact that the ''collision-dominated'' PM response dominates the Raman spectra at high temperatures (T $>$ $T_P$) and low fields ($H<$ 2T), making it difficult to measure the SF Raman response below $H$=2T for most temperatures. Consequently, in order to estimate $g_{eff}$ from our data, we first fit the field dependence of $\hbar \omega_o$ (for a particular temperature) in 
Figs.~\ref{BMPsumm} (a) and (b) using Eq. 7, solely for the purpose of verifying that the $H<$ 2T field range is in the linear-in-field regime. We then used the measured points falling in this linear-in-field regime to determine the slope of $\hbar \omega_o$ vs. $H$, and hence an experimental estimate of $g_{eff}$.  This estimate of $g_{eff}$ is quite close (a $\sim$ 10 $\%$ lower value) to that obtained if we simply estimate $g_{eff}$ from the slope of the Brillouin function fit in the linear-in-field regime.  A summary of our results is shown in Fig.~\ref{BMPsumm}(c).  Note that we estimate $g_{eff}$-values in EuB$_6$ and EuO of 75 (at T=35 K) and 60 (at T=90 K), respectively, which compare to values of $g_{eff}$=70 and 170 in $x$=0.01 and $x$=0.1 concentration Cd$_{1-x}$Mn$_x$Se, respectively.\cite{Heiman}

There is also an increase in the gaussian linewidth of the SF Raman response as a function of increasing field, implying that there is a corresponding increase in the distribution of polaron sizes with increasing magnetic field.  These trends are similar to those observed as a function of decreasing temperature at $H$=0, and are consistent with the field-dependence trends predicted by variational calculations.\cite{Guillaume}  These field-dependence studies are also consistent with a greater polaron stability in EuO as compared to EuB$_6$, since at $H$=8T we find that T$_P$ - T$_{\rm C}$($H$=0)$\sim$ 100 K in EuO as compared to T$_P$ - T$_{\rm C}$($H$=0) $\sim$ 40 K in EuB$_6$.  Again, this difference appears to reflect the greater static spin-disorder present in EuO.

Finally, it should be pointed out from Figs.~\ref{BMPsumm} (a) and (b) that even at higher temperatures for which there is no observable zero-field SF Raman response, the application of a magnetic field can induce the development of a SF Raman peak (see Fig.~\ref{BMP}). One interesting interpretation of this is that the application of a sufficiently high magnetic field in the spin-fluctuation dominated regime can induce the formation of magnetic polarons, presumably by reducing thermal fluctuations and by enhancing the magnetic susceptibility.

\subsubsection{Doping Dependence:  (Eu,La)B$_6$}
The (Eu,La)B$_6$ system are ideal materials in which to study the effects of doping on both the ferromagnetic metal transition and the properties of magnetic polarons.  The effects of La substitution are both to add one free electron and to increase spin- and charge-disorder through the removal of a Eu$^{2+}$ moment. The latter is evidenced by an increase in the room temperature resistivity with increased doping $x$, as shown in Fig.~\ref{susc resist}.  The effects of these different contributions can be separated to some extent by the application of a  magnetic field, which will decrease spin-disorder but will not affect charge-disorder.  

Information regarding the effects of a small amount of doping on the nature of the magnetic polaron can be gained by analyzing the behavior of the 
SF Raman response as a function of magnetic field at various temperatures and values of $x$.  Our results demonstrate that there is very little effect of a small amount of doping ($x \leq$ 0.01) on the magnetic polarons in (Eu,La)B$_6$, which is attributable to the highly local nature of the polarons.  Indeed, as shown in Figure~\ref{BMPsumm}, the SF Raman energies for the $x <$0.01 samples are only slightly lower (3-5 cm$^{-1}$) than for EuB$_6$, and Fig.~\ref{BMPsumm} (c) shows that the $g_{eff}$ values of (Eu,La)B$_6$ are rather insensitive to small amounts of doping.  Interestingly, the energies of the $x$=0.005 sample are slightly lower than those for the $x$=0.01 sample, possibly reflecting the differences observed in the plasma energy (see Fig.~\ref{optics}).  The $x$=0.01 sample also exhibits only slight effects of the higher carrier density on magnetic polaron formation:  for example, higher magnetic fields ($\sim$2 T) are required to induce a spin-flip Raman response at a given temperature in the $x$=0.01 sample compared to the $x$=0 sample.  It is also interesting to consider the higher doped samples, $x$ $\geq$ 0.03, in which no magnetic polarons are observed even at the lowest accessible temperatures and in magnetic fields up to 8 T.  This suggests that, at least at these fields, the carrier density (i.e., the carrier kinetic energy) is too high in these samples to permit magnetic polaron formation.  These data provide evidence that magnetic polarons form in low carrier density magnetic systems only if the carrier density is sufficiently low, and if the degree of spin-disorder is sufficient to localize charge, but not so high as to prevent local ferromagnetic alignment.  

\subsection{Ferromagnetic Metal Regime and the Metallization Transition}  
The transition from the magnetic polaron regime to the ferromagnetic metal regime in (Eu,La)B$_6$ and EuO is evident in the Raman scattering response as a crossover from an inelastic spin-flip Raman scattering response to a flat, magnetic-field-independent ''continuum'' response typical of that observed in other strongly-correlated metals (see Fig.~\ref{field near Tc}), 
including the high-T$_c$ cuprates and La$_{1-x}$Sr$_x$TiO$_4$.\cite{SrTiO3}  Note that the flat continuum response in the FM metal phase is associated with a strong electronic scattering intensity {\it above} background, and is therefore distinctly  different from the absence of electronic scattering intensity at low energies [see Fig.~\ref{field near Tc}] in the PM semimetal regime (Section III.A).  The continuum response observed in the FM metal regime is typically modelled using a collision-dominated response (Eq. 2) with a frequency-dependent scattering rate,\cite{SrTiO3} e.g., $\Gamma$($\omega$,T)=$\Gamma_0$(T) + a$\omega^2$ for the case of a Fermi liquid.\cite{fermi liquid}  While the microscopic origins of this response are not completely understood, it is generally believed to reflect a strongly-correlated metal phase dominated by strong inelastic carrier scattering from some broad spectrum of excitations.\cite{NyhusEuB6}  

The evolution of the Raman scattering response through the semiconductor-metal transitions in EuB$_6$ and EuO, illustrated in Figs.~\ref{bmp eub6} and~\ref{bmp euo} (a), provides important insight into the nature of this transition. In particular, this crossover provides a means of studying the manner in which magnetic polarons evolve through the semiconductor-metal transition, and into the ferromagnetic metal phase, in these low T$_{\rm C}$ systems. Significantly, the evolution of the SF Raman response into the ferromagnetic phase in these materials can be most sensitively studied as a function of increasing magnetic field: physically, a field-induced transition is associated with the reduction of spin-disorder, and the consequent systematic increase in the carrier localization length, with increasing magnetic field. Such a field-induced transition is illustrated in Fig.~\ref{field near Tc}  for EuB$_6$, and is clearly similar to temperature-dependent transitions in EuB$_6$ and EuO shown in Figs.~\ref{bmp eub6} and~\ref{bmp euo} (a). Fig.~\ref{field near Tc}  provides evidence that the magnetic polaron regime continuously evolves into the ferromagnetic metal regime with increasing field (and decreasing temperature). Specifically, the gaussian linewidth of the SF Raman response rapidly grows with increasing magnetic field as the transition is approached (see inset, Fig.~\ref{field near Tc}), suggesting that there is a systematic increase in the distribution of polaron sizes,\cite{Calderon,Guillaume} and the polaron response continuously evolves into the flat 'continuum' response associated with the FM metallic phase. Notably, however, there is one important difference in the manner in which polarons evolve in EuB$_6$ and EuO:  in EuB$_6$, the delocalization of the polaron appears to be complete, resulting in a homogeneous ferromagnetic state; this is supported by the absence of a polaronic response, and the complete field-independence of the 'continuum' response, below T$_{\rm C}$. By contrast, in EuO there is evidence that the polaron response persists to some extent below T$_{\rm C}$, even in magnetic fields up to $H$ = 8 T, suggesting that the ferromagnetic metal phase may be inhomogeneous in EuO. A possible reason for this difference is the greater amount of static spin-disorder present in EuO compared to EuB$_6$ (see Section III.C.1), which stabilizes small polarons to some extent even in the presence of a larger carrier concentration in the FM metal phase. Notably, a similar persistence of small polarons in the ferromagnetic phase is believed to occur also in the colossal magnetoresistance phase of the manganites, again due to the large amount of intrinsic disorder affiliated with local Jahn-Teller and breathing mode lattice distortions that persist below T$_{\rm C}$.\cite{Yoon}

\section{Conclusions}
EuO has long been studied as a model material in which to study paramagnetic-semiconductor to ferromagnetic-metal transitions, while EuB$_6$ has only recently attracted interest due to its many similarities to the manganese perovskites.  In this study, we provide direct spectroscopic evidence that the PM-semiconductor to FM-metal transitions in EuO and EuB$_6$ are preceded by the formation of magnetic polarons.  Also, we show that in the PM phase of doped magnetic systems, the spin dynamics can be effectively studied with light scattering via 'collision-dominated' electronic Raman scattering involving spin fluctuation scattering.  Importantly, Raman scattering allows us to probe previously unexplored regions of the complex phase regimes of the magnetic semiconductors EuO and EuB$_6$, and to study the effects of temperature, magnetic field, and doping on their metal-insulator transitions, magnetic polarons, and spin dynamics.  We show that a magnetic field can be used to tune the spin-disorder and magnetic susceptibility so that conditions will favor magnetic polaron formation, and drive the FM metal transition.  We further show that low levels of doping, $x \leq$ 0.01, and isotopic substition, have very little effect on the magnetic polaron energies, consistent with the local, and pure spin, nature of the polarons.  At higher doping levels, $x \geq$ 0.03, however, the carrier kinetic energies are sufficiently high that magnetic polarons are not observed at temperatures down to 18 K and magnetic fields as high as 8 T.  The effects of carrier mobility and spin disorder on magnetic polaron formation and stability were demonstrated by juxtaposing the results observed for (Eu,La)B$_6$ with those observed for the more intrinsically disordered EuO system.  In particular, low carrier mobility was seen to play a key role in stabilizing the magnetic polarons both above and below T$_{\rm C}$.  Finally, by demonstrating its efficacy for elucidating the complex nature of EuO and EuB$_6$, we demonstrate in this study the unique abilities of Raman scattering for probing the complex magnetic and electronic phenomena of doped magnetic systems, including critical spin dynamics, magnetic polaron formation, and metal/semiconductor transitions.

\section{ACKNOWLEDGMENTS}
This work was supported by the Department of Energy under Grant No. DEFG02-96ER45439 (C.S.S. and S.L.C.), by the National Science Foundation through DMR97-00716 (C.S.S. and S.L.C.), and by the NHMFL through Grant No. NSF DMR90-16241 (Z.F. and D.P.Y.).  We thank Hsiang-Lin Liu for making the room temperature reflectivity measurements, S.Y. Yoon for initial measurements of the Eu$^{10}$B$_6$ sample, and I.S.Yang for help in data taking.  We also acknowledge use of the FS-MRL Laser Facility for some of the measurements.

\begin{figure}
\caption{(Top) Magnetic susceptibility $\chi$ (emu/gm) vs. Temperature. (Bottom)  Resistivity($\mu\Omega$-cm) vs. Temperature.  The paramagnetic semimetal to ferromagnetic metal transition temperatures are 15.4 K, 13.7 K, and 12.45 K for the EuB$_6$, x=0.005, and x=0.01 samples, respectively, as determined by the position of the peak in d$\rho$/dT.\cite{SullowMag}} 
\label{susc resist} 
\end{figure}

\begin{figure}
\caption{(Top)  T=293 K optical conductivity, and (bottom) dielectric loss function, for Eu$_{1-x}$La$_x$B$_6$.  The dramatic shifts occur between the x=0.01 and x=0.03 samples.}  
\label{optics} 
\end{figure}

\begin{figure}
\caption{Qualitative Phase Diagram of the Eu$_{1-x}$La$_x$B$_6$ system.  Squares with crosses roughly denote our samples at the lowest obtainable temperature of 18 K.}  
\label{phasediagram} 
\end{figure}

\begin{figure}
\caption{Raman spectra of Eu$^{11}$B$_6$ showing the three Raman active phonons at 
762 cm$^{-1}$ (T$_{2g}$), 1098 cm$^{-1}$ (E$_g$), and 1238 cm$^{-1}$ (A$_{1g}$), and the low-energy diffusive scattering response.  The dashed line is a fit of Eq. 2.}
\label{EuB6 300K}
\end{figure}

\begin{figure}
\caption{Raman response of Eu$_{1-x}$La$_x$B$_6$ at T=300 K.  Inset demonstrates the relationship between the integrated spectral weight I($\omega$), and $\omega_p^2$ as a function of $x$.}
\label{300k}  
\end{figure}

\begin{figure}
\caption{$\Gamma$ (left axis) and T$\chi$ (right axis) scale for temperatures approaching the transition temperature, 15.4 K, for EuB$_6$.}
\label{gamma chi}
\end{figure}

\begin{figure}
\caption{(a)  Raman spectra of EuO at temperatures above T$_{\rm C}$ for $H=0$.  The peak that develops at 80 K is $H$=0 spin-flip Raman scattering associated with the development of magnetic polarons, discussed in Section III.B.1.  (b) Integrated spectral weight I($\omega$) vs. Temperature for EuO.}
\label{EuO 1}
\end{figure}

\begin{figure}
\caption{Three distinct Raman scattering regions in EuB$_6$ for $H=0$, a spin fluctuation regime for 65 $>$ T $>$ 30 K, magnetic polaron (MP) regime for 30 $>$ T $>$ 15 K, and a ferromagnetic metal phase below T=15 K.  Inset shows behavior of $\Gamma$ as a function of temperature for EuB$_6$ and defines four distinct material phases. 1) Paramagnetic semimetal, 2) spin fluctuation, 3) MP, and 4) ferromagnetic metal.}  
\label{bmp eub6} 
\end{figure}

\begin{figure}
\caption{(a)  Raman spectra of EuO showing the evolution of the $H$=0 SF Raman peak with decreasing temperature towards T$_{\rm C}$.  The inelastic gaussian peak that develops around 35 cm$^{-1}$ at 80 K is $H$=0 spin-flip Raman scattering associated with the presence of magnetic polarons.  Note that the magnetic polaron persists even at temperatures below T$_{\rm C}$.  Inset shows the temperature dependence of the inelastic peak as a function of temperature. (b)  Symmetry dependence of the spin-flip Raman spectra, indicating distinctive E$_i \perp$ E$_s$ symmetry.  The dashed line is a fit to a gaussian function.}  
\label{bmp euo} 
\end{figure}

\begin{figure}
\caption{(a) T=20 K Raman response for EuB$_6$ in various magnetic 
fields.  The SF Raman peak due to magnetic polarons becomes observable near 100 cm$^{-1}$ at $H \sim$8T at this temperature. (b) T=35 K Raman response for the $x$=0.01 sample in various magnetic fields.  (c)  T=100 K Raman response for EuO in various magnetic fields.  The SF Raman 
response becomes observable near 60 cm$^{-1}$ at $H \sim~$8T at this temperature.} 
\label{BMP}
\end{figure}

\begin{figure}
\caption{Field dependence of the field-dependent peak, $\hbar \omega_0$, at various temperatures for (a)  Eu$_{1-x}$La$_x$B$_6$, and (b)  EuO.  The lines are drawn as guides to the eye.  Note the presence of a SF Raman peak in both EuB$_6$ and EuO, even in the absence of a magnetic field.  (c)  ($\delta \hbar \omega_o/ \delta H$) and (${1 \over \mu_B}$)($\delta \hbar \omega_o/ \delta H$) vs. Temperature for (Eu,La)B$_6$ and EuO.  The solid lines are guides to the eye.  Also shown is data for Cd$_{0.99}$Mn$_{0.01}$Se, from ref. 30.}
\label{BMPsumm}
\end{figure}

\begin{figure}
\caption{The optical modes of Eu$^{10}$B$_6$ are shifted by 5$\%$ from the energies of the Eu$^{11}$B$_6$ system.  The inset shows the SF Raman response for EuB$_6$ and the Boron isotope substituted sample at T=20 K and $H$=8T.}
\label{boron10}
\end{figure}

\begin{figure}
\caption{Raman spectra of EuB$_6$ taken at T=18 K and various magnetic fields.  The inset shows the width of the Gaussian profile for the SF Raman response, indicating an increase in the average magnetic polaron size with increasing magnetic field.  The average size of the polaron increases and diverges near the transition field and temperature.}
\label{field near Tc}
\end{figure}

\begin{references}
\bibitem{t-tc}  The typical $T$ dependence of the Brillouin function needs to be modified to account for the ground state of the material, in the same fashion that the magnetic susceptibility is modified.  For antiferromagnetic materials $T$ is replaced by $T+T_{\rm AF}$, where $T_{\rm AF}$ accounts for antiferromagnetic interactions.  For ferromagnetic materials, T$^*$ reflects modifications of $T$ due to the Curie temperature.  See refs. 16,17, and 30.
\bibitem{droplets}  Note also that this magnetic polaron analysis ignores the possibility that ferromagnetic clusters involving more than one carrier, i.e., magnetic ''droplets'' (see ref. 10), also develop above T$_{\rm C}$ in these materials.
\end{references}
\end{document}